\documentclass[aps,prd,reprint,groupedaddress,nofootinbib]{revtex4-2}

\usepackage{fullpage,epsf,amssymb,amsthm,amsfonts,amsmath,latexsym,array,graphicx,extarrows,mathtools,mathstyle,mathrsfs,slashed,amsbsy,csquotes,accents,enumitem}

\begin{document}
    
\title{Goldstone bosons at non-zero temperature}  

\author{Peter Lowdon}
\author{Owe Philipsen}
\affiliation{ITP, Goethe-Universit\"{a}t Frankfurt am Main, Max-von-Laue-Str. 1, 60438 Frankfurt am Main, Germany}

\begin{abstract}
Spontaneous symmetry breaking in quantum field theories at non-zero temperature still holds fundamental open questions, in particular what happens to vacuum Goldstone bosons when the temperature is increased. By investigating a complex scalar field theory on the lattice we demonstrate that Goldstone bosons at non-zero temperature behave like screened massless particle-like excitations, so-called thermoparticles, which continue to exist even in the symmetry-restored phase of the theory. We provide non-perturbative evidence for the functional form of the Goldstone mode's dissipative behaviour, and determine its corresponding spectral properties. Since the persistence of thermal Goldstone modes within symmetry-restored phases is predicted to be a model-independent characteristic, this has fundamental consequences for systems in which continuous symmetries are restored at high temperatures.
\end{abstract}

\maketitle

\noindent
{\bf{\textit{Introduction}}} \\
The spontaneous breaking of continuous symmetries plays a fundamental role for many physical systems. In the relativistic regime, Goldstone's theorem~\cite{Goldstone:1962es,Kastler:1966wdu} states that the spontaneous breaking of a continuous symmetry implies the existence of massless bosons. Although the theoretical and experimental consequences of this theorem are well-understood at zero temperature, relatively little is known from first principles about systems at non-zero temperature, and in particular those which undergo phase transitions at high temperatures~\cite{Kirzhnits:1972ut,Dolan:1973qd,Weinberg:1974hy}. Understanding these characteristics is important for describing numerous phenomena, including phase transitions in the early universe, or the phase diagram of nuclear matter. \\
\indent
For relativistic systems at zero-temperature, Goldstone bosons leave distinct signatures on the correlation functions of the theory, in particular the appearance of $\delta(p^{2})$ singularities in the Fourier transform of the two-point function $\langle [j_{0}(x), A(y)]\rangle$, where $j_{\mu}$ is the conserved current associated with the symmetry, and $A$ is a local field whose transformation under the symmetry has a non-vanishing vacuum expectation value, $\langle \delta A \rangle \neq 0$~\cite{Strocchi:2021abr}. For non-zero temperatures $T=1/\beta>0$ the boost invariance of the system is lost, and it is no longer clear that stable massless Goldstone modes will exist. Nevertheless, the Fourier transform of the thermal commutator $\langle [j_{0}(x), A(y)]\rangle_{\beta}$ continues to contain a zero-energy singularity $\delta(\omega)$ in the limit $\vec{p}\rightarrow 0$, a Goldstone quasi-particle~\cite{Strocchi:2021abr}. In Ref.~\cite{Bros:1996kd} the authors made a significant breakthrough by further demonstrating that Goldstone modes for $T>0$ have the structure of distinct particle-like excitations. To reach this conclusion the authors considered finite-temperature systems which satisfy the fundamental non-perturbative QFT constraints proposed in Ref.~\cite{Bros:1996mw}, namely that one has local causal fields\footnote{For a complex scalar theory causality means that the fields satisfy: $\left[\phi(x),\phi^{\dagger}(y)\right]=0$ for $(x-y)^{2}<0$, which guarantees that space-like separated measurements commute with one another and are therefore causal~\cite{Streater:1989vi,Haag:1992hx,Bogolyubov:1990kw}.} which transform covariantly under spacetime translations and spatial rotations, and a thermal ground state which is invariant with respect to these transformations. Under these conditions one can demonstrate that the thermal expectation values of field commutators satisfy a non-perturbative spectral representation~\cite{Bros:1992ey,Bros:1996mw}. In particular, for a complex scalar field $\phi(x)$ at non-zero temperature the spectral function $\rho(\omega,\vec{p})$, defined as the Fourier transform of the thermal commutator $\langle[\phi(x),\phi^{\dagger}(0)]\rangle_{\beta}$, has the following general form:
\begin{align}
\rho(\omega,\vec{p}) &=  \int_{0}^{\infty} \!\! ds  \int \! \frac{d^{3}\vec{u}}{(2\pi)^{2}} \  \epsilon(\omega) \nonumber \\
& \quad\quad \times  \delta\!\left(\omega^{2} - (\vec{p}-\vec{u})^{2} - s \right) \widetilde{D}_{\beta}(\vec{u},s), 
\label{spec_rep}
\end{align}
where $\epsilon(\omega)$ is the sign function. This represents the $T>0$ generalisation of the well-known K\"{a}ll\'{e}n-Lehmann representation that exists for QFTs at $T=0$~\cite{Kallen:1952zz,Lehmann:1954xi}. An important implication of Eq.~\eqref{spec_rep} is that the behaviour of the spectral function is fixed by the thermal spectral density $\widetilde{D}_{\beta}(\vec{u},s)$, and so its properties hold the key for determining the type of excitations that can exist, and how they are modified by changes in $T$. Since the Fourier transform of $\langle [j_{0}(x), A(y)]\rangle_{\beta}$ also satisfies an analogous representation~\cite{Bros:1996kd}, it follows from Goldstone's theorem that the position space thermal spectral density must satisfy the condition: $D_{\beta}(\vec{x},s) \rightarrow \delta(s)$ for $T\rightarrow 0$, since then Eq.~\eqref{spec_rep} implies the existence of a distinct massless component $\delta(p^{2})$. In Ref.~\cite{Bros:1996kd} it is explicitly demonstrated that the vacuum Goldstone singularity in $D_{\beta}(\vec{x},s)$ persists for $T>0$, even if the symmetry is restored at high temperatures. In particular, this means that the thermal spectral density contains a distinguished Goldstone contribution of the form
\begin{align}
D_{\beta}^{G}(\vec{x},s) = D_{\beta}^{G}(\vec{x})\delta(s).   
\label{D_G}
\end{align}
When $D_{\beta}^{G}(\vec{x})$ is non-trivial this causes the stable massless Goldstone peak in the spectral function at $p^{2}=0$ to become broadened, which describes the dissipative effects that the Goldstone boson experiences as it moves through the thermal medium. Since $D_{\beta}^{G}(\vec{x})$ also reduces the amplitude of the Goldstone propagator, it represents a thermal damping factor. The particle-like structure described by Eq.~\eqref{D_G} is the massless realisation of a general proposition $D_{\beta}(\vec{x},s) = D_{\beta}(\vec{x})\delta(s-m^{2})$, first put forward in Ref.~\cite{Bros:1992ey} for how stable vacuum particle states with mass $m$ should behave when $T>0$. These were later referred to as \textit{thermoparticles} in order to draw a distinction between other types of thermal excitations such as quasiparticles~\cite{Buchholz:1993kp}. Evidence for the existence of massive thermoparticles has since been found in scalar theories~\cite{Lowdon:2024atn}, as well as more complex theories such as quantum chromodynamics (QCD)~\cite{Lowdon:2022xcl,Bala:2023iqu}. \\

\noindent
{\bf{\textit{Signatures in Euclidean correlation functions}}} \\
If Goldstone bosons behave like massless thermoparticles when $T>0$, these excitations will leave distinct signatures on the correlation functions of the theory. Understanding the impact this has on the behaviour of Euclidean correlation functions is important since many of the non-perturbative techniques for studying thermal correlation functions, such as lattice calculations, are restricted to or optimised for calculations in imaginary time. In this work we will focus on QFTs involving complex scalar fields, where the real-time thermal two-point function is defined $\mathcal{W}(x_{0},\vec{x})=\langle\phi(x_{0},\vec{x})\phi^{\dagger}(0)\rangle_{\beta}$. Due to the spectral representation in Eq.~\eqref{spec_rep}, and the condition of thermal equilibrium\footnote{The Kubo-Martin-Schwinger (KMS) condition~\cite{Haag:1967sg}: $\mathcal{W}(x_{0},\vec{x}) = \mathcal{W}(-x_{0}-i\beta,-\vec{x})$ defines the notion of thermal equilibrium, and implies the following connection between the momentum-space two-point function and spectral function: $\widetilde{\mathcal{W}}(\omega,\vec{p})= \rho(\omega,\vec{p})(1-e^{-\beta\omega})^{-1}$.}, it follows that the two-point function also possesses a spectral representation
\begin{align}
\mathcal{W}(x_{0},\vec{x}) = \int_{0}^{\infty} \! ds \ \mathcal{W}_{\beta}^{(s)}(x_{0},\vec{x}) \, D_{\beta}(\vec{x},s),
\label{W_rep}
\end{align}
where $\mathcal{W}^{(s)}(x_{0},\vec{x})$ is the thermal two-point function for a free particle\footnote{This two-point function has the form: $\mathcal{W}^{(s)}(x_{0},\vec{x}) = \int \frac{d^{4}p}{(2\pi)^{4}}e^{-ip \cdot x} \, 2\pi\epsilon(\omega) \, \delta\!\left(\omega^{2} - |\vec{p}|^{2} - s \right)(1-e^{-\beta\omega})^{-1}$.} with mass $\sqrt{s}$. For the Euclidean two-point function $C(\tau,\vec{x})=\langle\phi(\tau,\vec{x})\phi^{\dagger}(0)\rangle_{\beta}$, understanding the spatial variation of $C(\tau=0,\vec{x})$ amounts to determining $\mathcal{W}(x_{0}=0,\vec{x})$. Equation~\eqref{W_rep} therefore implies that the Goldstone mode in Eq.~\eqref{D_G} gives the following discrete contribution to the spatial Euclidean two-point function:  
\begin{align}
C^{G}(0,\vec{x}) = \frac{\coth\left(\frac{\pi |\vec{x}|}{\beta} \right)}{4\pi \beta |\vec{x}| }D_{\beta}^{G}(\vec{x}).
\label{Goldstone_T}
\end{align}
Since $D_{\beta}^{G}(\vec{x}) \xrightarrow[]{T \rightarrow 0} \alpha_{0}$, with $\alpha_{0}$ a constant, the Goldstone contribution to the two-point function reduces to that of a massless vacuum particle in this limit
\begin{align}
C^{G}(0,\vec{x}) \xlongrightarrow[]{T\rightarrow 0}\frac{\alpha_{0}}{4\pi^{2}|\vec{x}|^{2}}.
\label{Goldstone_T0}
\end{align}
Another Euclidean correlation function of particular relevance is the spatial screening correlator along the $z$-axis, which has the form
\begin{align}
C(z) &= \int  \!dx \, dy \, d\tau \, C(\tau,\vec{x})\nonumber \\ 
&= \frac{1}{2}\int_{0}^{\infty} \! ds \int^{\infty}_{|z|} \! dR \ e^{-R\sqrt{s}} D_{\beta}(R,s),
\label{C_int}
\end{align}
where the representation in the second line follows from Eq.~\eqref{spec_rep}~\cite{Lowdon:2022xcl}. Here one defines: $D_{\beta}(R,s) = D_{\beta}(|\vec{x}|=R,s)$, which makes use of the fact that the thermal spectral density only depends on $|\vec{x}|$ due to rotational invariance. In this case the Goldstone mode gives the contribution
\begin{align}
C^{G}(z) = \frac{1}{2} \int^{\infty}_{|z|} \! dR \  D_{\beta}^{G}(R).
\label{CG_int}
\end{align}
Now that we understand the impact of the thermal Goldstone modes derived in Ref.~\cite{Bros:1996kd} on Euclidean correlation functions, we will investigate the presence of these modes in lattice data. \\    

\noindent
{\bf{\textit{Analysis of $\mathrm{U}(1)$ scalar lattice field theory}}} \\
The simplest QFT model for which a continuous symmetry is spontaneously broken in vacuum and restored at high temperatures is the $\mathrm{U}(1)$ complex scalar field theory. In the broken phase of the theory at $T=0$, the scalar field has a non-vanishing vacuum expectation value $|v|^{2}=\langle \phi \rangle\langle \phi^{\dagger} \rangle$, and
the model contains a massless Goldstone mode and a resonance-like $\sigma$ mode, which is unstable on account of its decay into Goldstone bosons. The model is expected to undergo a second-order phase transition at some critical temperature $T_{c}$, above which $|v|^{2}=0$ and the global $\mathrm{U}(1)$ symmetry is restored\footnote{For discussions regarding the perturbative characteristics of this model see~\cite{Kapusta:2006pm,Pilaftsis:2013xna} and the references within.}. For our analysis we work with the following lattice discretisation of the action
\begin{align}
S &= a^{4} \! \sum_{x \in \Lambda_{a}} \Bigg[ \sum_{\mu} \left( \frac{1}{2} \Delta_{\mu}^{f} \phi^{*}(x)\Delta_{\mu}^{f}\phi(x)\right) \nonumber \\
& \quad\quad\quad + \frac{m_{0}^{2}}{2}\phi^{*}(x)\phi(x) + \frac{g_0}{4!}\left(\phi^{*}(x)\phi(x)\right)^{2} \Bigg],
\label{Latt_action}
\end{align}
where $\Delta_{\mu}^{f}$ is the lattice forward derivative, and $a>0$ is the lattice spacing. In order to avoid the potential triviality of the model we keep the lattice spacing fixed throughout, so that it is sufficient to consider non-renormalised correlation functions, while $T= (aN_{\tau})^{-1}$ can be varied in discrete steps only. Whether one is in the broken phase or not depends on the specific value of the bare parameters $m_{0}$ and $g_{0}$. For fixed $g_{0}$, it turns out that $(am_{0})^{2}$ is necessarily negative in the broken phase, and hence $am_{0}$ must be a pure imaginary number. We further require a sufficiently fine and large lattice to ensure that the lattice temperature covers both the symmetry-broken and restored phases for large and small values of $N_{\tau}$, respectively. Having scanned a range of bare parameter values, we found that $(am_{0},g_{0})=(0.297 i,0.85)$ was a good choice. In the vacuum broken symmetry phase the expectation value $|v|$ sets the physical scale, and separates long $|\vec{x}||v|>1$ and short $|\vec{x}||v|<1$ distances. As we shall see, our choice of parameters results in a short distance range $|\vec{x}| \lesssim 11 a$, and a UV cutoff $\Lambda/|v|=\pi/(a|v|) \approx 36$, such that cutoff effects should be very small for lattice correlators beyond a few lattice spacings. 
  
Determining the phase of any lattice model numerically is a non-trivial task. Spontaneous symmetry breaking cannot occur in a finite spatial volume $V$, and thus requires a $V\rightarrow \infty$ extrapolation of the lattice correlation functions. On a spatially symmetric space-time volume $L^{3}\times L_\tau$, where $L_{\tau}= aN_{\tau}$ and $L = aN_{s}$, the finite spatial-volume Euclidean two-point function $C_{L}(\tau,\vec{x}) = \langle \phi(\tau,\vec{x}) \phi^{\dagger}(0)\rangle_{L}$ satisfies the condition
\begin{align}
\lim_{L\rightarrow \infty}C_{L}(\tau,\vec{x}) \xrightarrow[]{|\vec{x}\hspace{0.3mm}|\rightarrow \infty} |v|^{2}.
\end{align}
There are different approaches for extracting $|v|^{2}$ from lattice data~\cite{Neuberger:1987fd,Neuberger:1987zz}, but a simple way is to consider the limit of correlators evaluated at their largest spatial extent
\begin{align}
|v|^{2} = \lim_{L\rightarrow \infty}  C_{L}(0,|\vec{x}|=L/2).
\end{align}  
An estimate of $|v|^{2}$ can then be obtained by extrapolating the $L\rightarrow \infty$ behaviour using a range of correlators at sufficiently large values of $N_{s}$.

\begin{figure}[t!]    
\includegraphics[width=0.47\textwidth]{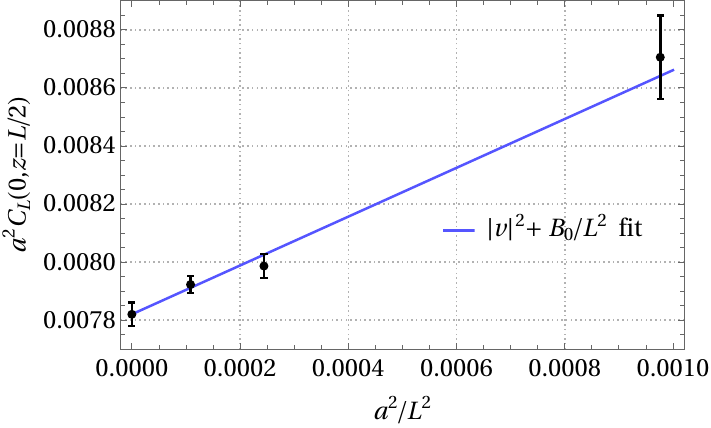} 
\caption{Infinite-volume extrapolation of $a^{2}C_{L}(0,L/2)$.}
\label{v_extrap}
\end{figure}

For $N_{\tau}=32$ we expect the system to be in the low-temperature, vacuum-like broken phase, where the Goldstone mode dominates the behaviour of $C_{L}(0,\vec{x})$, and the connected correlator should approach the massless particle form in Eq.~\eqref{Goldstone_T0} for sufficiently large volumes. To test this hypothesis, we performed fits of lattice correlator data for different volumes ($N_{s}= 32, 64, 96$) with the finite-volume functional form
\begin{align}
C_{L}(0,z) = c_{0} + b_{0} \left[ \tfrac{1}{z^{2}} + \left\{ z \rightarrow (L-z)\right\} \right], 
\label{vac_fit}
\end{align}
where $C_{L}(0,z)=  C_{L}(0,|\vec{x}|=z)$, and the final term accounts for the spatially periodic boundary conditions of our lattices. The fits were performed over a spatial range $[z_{\text{min}},L/2]$, where $z=L/2$ is the maximal lattice correlator extent, and $z_{\text{min}}<L/2$ was varied in order to assess the fit stability. Keeping $z_{\text{min}}/a>5$ in order to avoid cutoff effects, we found that the continuum form Eq.~\eqref{vac_fit} provided a very good description of the data across large ranges for each of the volumes considered. To estimate $|v|^{2}$ we therefore fit $C_{L}(0,z=L/2)$ to the functional form $|v|^{2} + B_{0}/L^{2}$, as shown in Fig.~\ref{v_extrap}, resulting in the non-zero infinite-volume extrapolation
\begin{align}
a^{2}|v|^{2} = 0.00782(4).
\end{align}
The ratio $T/|v|=1/(N_{\tau}a|v|)$ provides a qualitative measure for the temperature of the system, which in this case $T/|v|\approx 0.35$ confirms that it is in a cold vacuum-like state\footnote{Strictly speaking, $|v|^{2}$ also requires an infinite $N_{\tau}$ extrapolation in order to represent the true vacuum value. However, the precise value is not important for our qualitative analysis, only that $|v|^{2}$ is significantly non-vanishing.}. Overall, these results demonstrate that the $N_{\tau}=32$ data is in the broken phase in the infinite-volume limit, and that the scalar correlation function is dominated by a massless vacuum-like Goldstone state, as is the case in the broken phase of the $\mathrm{O}(4)$ model~\cite{Hasenfratz:1987eh,Hasenfratz:1988kr,Hasenfratz:1989ux,Hasenfratz:1990fu,Gockeler:1992zj}. 

\begin{figure}[t]  
\includegraphics[width=0.47\textwidth]{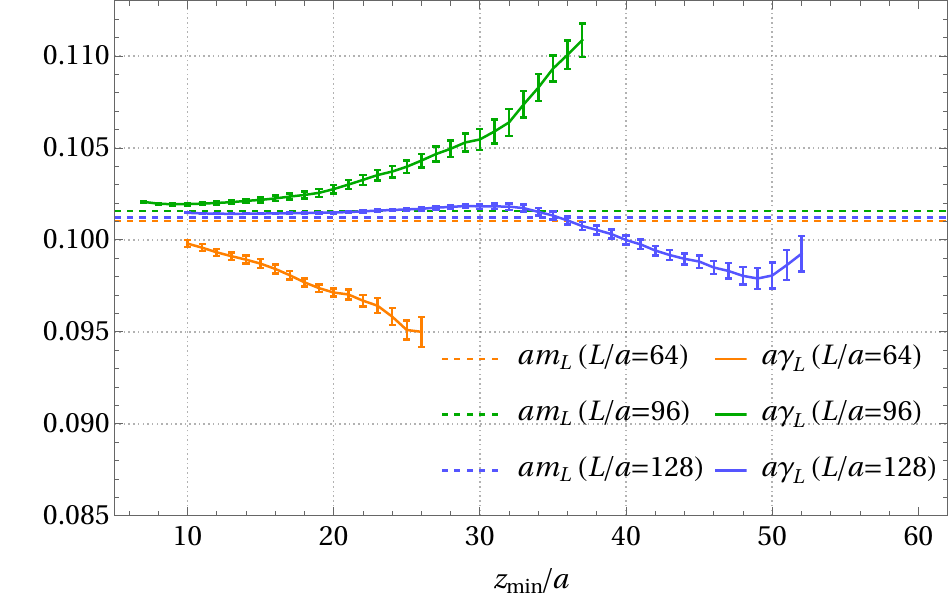}
\caption{Fitted values of $\gamma_{L}$ as a function of $z_{\text{min}}$ for different volumes. The horizontal dashed lines indicate the best-fit values of $m_{L}$ extracted from the spatial screening correlator data.}
\label{gammaL:fit}
\end{figure}

By contrast, for $N_{\tau}=2$ we have $T/|v|\approx 5.65$, which indicates that the system is in the symmetry-restored phase. As discussed above, the Goldstone mode can still exist in this phase even though the symmetry is no longer broken, but it has the structure of a massless thermoparticle~\cite{Bros:1996kd}. If this mode provides a dominant contribution, then $C(0,\vec{x})$ will take the form of Eq.~\eqref{Goldstone_T}. By investigating the properties of the spatial screening correlator $C(z)$, one can use Eq.~\eqref{CG_int} to determine the qualitative structure of the Goldstone damping factor $D_{\beta}^{G}(\vec{x})$. For this purpose, we performed fits to the lattice spatial screening correlator data with the finite-volume single-exponential ansatz
\begin{align}
C_{L}(z) = d_{L} \left[ e^{-m_{L} z} +  \left\{ z \rightarrow (L-z)\right\} \right]. \label{Cz_L}  
\end{align}
We found that Eq.~\eqref{Cz_L} provides an excellent description of the data across the full range $[0,L/2]$ for each of the volumes considered ($N_{s}=64, 96, 128$), with $\chi^{2}/\text{d.o.f.} \lesssim 1$ in all cases. The fit values obtained for $m_{L}$ are plotted as dashed lines in Fig.~\ref{gammaL:fit}, and have a relatively weak volume dependence. This strongly suggests that the theory is indeed in the symmetry-restored phase in the infinite-volume limit, and that due to Eq.~\eqref{CG_int} the Goldstone mode must have an exponential damping factor: $D_{\beta}^{G}(\vec{x}) = \alpha \, e^{-\gamma |\vec{x}|}$, and hence the spatial two-point function has the form 
\begin{align}
C^{G}(0,\vec{x})  =  \frac{\coth\left(\frac{\pi |\vec{x}|}{\beta} \right)}{4\pi \beta |\vec{x}| }\alpha \, e^{-\gamma |\vec{x}|}.
\label{CG}
\end{align} 
To test the consistency of this conclusion we fit the lattice spatial two-point function data using the finite-volume ansatz
\begin{align}
\!\!\! C_{L}(0,z) \! =  b_{L} \! \left[\tfrac{\coth\left(\frac{\pi z}{\beta} \right)}{z} e^{-\gamma_{L} z} + \!\left\{z \rightarrow (L-z)\right\} \! \right] \!. 
\label{C_L}   
\end{align}
The fits were performed over a range $[z_{\text{min}},L/2]$, and the quality and stability of the fits were assessed by computing their sensitivity to $z_{\text{min}}$. We found that Eq.~\eqref{C_L} described the data increasingly better for larger volumes, as can be seen in Fig.~\ref{gammaL:fit} by the improved stabilisation of the fit values for $\gamma_{L}$ as a function of $z_{\text{min}}$. In this case, the volume dependence of the fits was significantly more pronounced than for the spatial screening correlator. As in the broken phase, the $C_{L}(0,z)$ fits were restricted to $z_{\text{min}}$ values larger than a few lattice spacings in order to avoid cutoff effects. Since $\coth(\pi z/\beta)z^{-1}\approx z^{-1}$ for almost all values of $z$, to further assess the robustness of Eq.~\eqref{C_L} we also performed fits using a range of parametrisations of the form: $B_{L} \left[ z^{-n} e^{-\Gamma_{L} z} + \{ z \rightarrow (L-z) \} \right]$ with $n \neq 1$, and also without exponential factors. In all of these cases we found that the fit parameter values were highly sensitive to $z_{\text{min}}$, and hence did not provide a good description of the data. If lattice cutoff effects are small, then a highly non-trivial test that the Goldstone mode has a massless thermoparticle structure is that the screening mass $m_{L}$ and damping factor exponent $\gamma_{L}$ must converge in the infinite-volume limit: 
\begin{align}
\lim_{L \rightarrow \infty} \gamma_{L} = \lim_{L \rightarrow \infty} m_{L}  = \gamma.
\label{gammaL}
\end{align}
In Fig.~\ref{gammaL:fit} we plot the values of $\gamma_{L}$ obtained in the fit range $[z_{\text{min}},L/2]$ for which $\chi^{2}/\text{d.o.f.} \lesssim 1$, and where the fit errors are less than 1\%. One can see that the $m_{L}$ and $\gamma_{L}$ values increasingly approach one another for larger volumes, which strongly indicates that the condition 
in Eq.~(\ref{gammaL}) is satisfied. 

Using the representation in Eq.~\eqref{spec_rep} together with the damping factor $D_{\beta}^{G}(\vec{x})$, one can compute the corresponding spectral function of the Goldstone boson $\rho_{G}(\omega,\vec{p})$, which has the explicit form
\begin{align}
\rho_{G}(\omega,\vec{p}) = \frac{4\alpha \, \omega \gamma }{(\omega^{2}-|\vec{p}|^{2}-\gamma^{2})^{2} + 4\omega^{2}\gamma^{2}}.
\label{rhoG}
\end{align}

\begin{figure}[t]  
\includegraphics[width=0.49\textwidth]{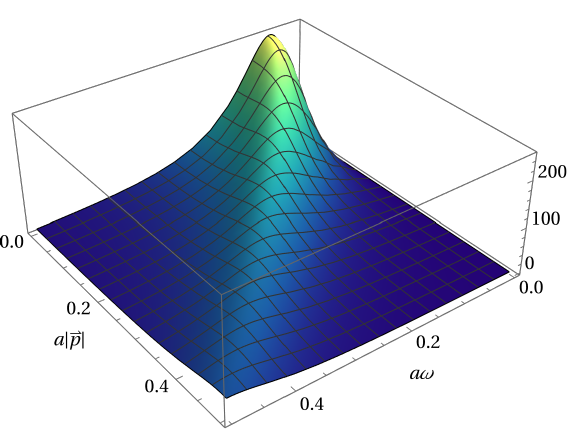}
\caption{Goldstone spectral function $\rho_{G}(a\omega,a|\vec{p}|)/a^{2}$ on the largest hot lattice, $N_{s}=128$ at $N_{\tau}=2$.}
\label{spectral:G}
\end{figure}

In Fig.~\ref{spectral:G} we plot $\rho_{G}$ using the results obtained from the fits on the largest lattice volume $N_{s}=128$. 
In contrast to the vacuum-like case at $N_{\tau}=32$, where the Goldstone boson has the structure of a massless particle state with spectral function proportional to $\delta(p^{2})$, at $N_{\tau}=2$ the Goldstone mode experiences appreciable dissipative effects due to its interactions with the thermal medium, resulting in a spectral peak broadened around the vacuum singularity $p^{2}=0$. The Goldstone spectral function has similar properties to that of a vacuum resonance state, except that for $\rho_{G}$ the width arises from purely collisional processes, as opposed to mixing effects brought about by the intrinsic instability of the state.  \\

\noindent
{\bf{\textit{Physical implications}}} \\
In this work we have focussed on the $\mathrm{U}(1)$ complex scalar theory, which confirms the fundamental prediction of Ref.~\cite{Bros:1996kd} that the Goldstone mode can persist in the symmetry-restored phase, and has a massless thermoparticle structure. This prediction characterises a model-independent property of Goldstone modes in QFTs at finite temperature, and therefore applies to all such systems in which a continuous symmetry is restored by thermal effects. Two particularly consequential implications are:
\begin{itemize}[leftmargin=*]

\item In the non-relativistic limit, any Goldstone mode which continues to exist in the low-temperature phase of a system should also leave distinct observable signatures above the corresponding critical temperature $T_{c}$. This applies, for example, to zero or low-density condensed matter systems, and could be used to determine the $T>T_{c}$ spectral properties of thermal Goldstone modes, such as spin waves in ferromagnets, or Landau phonons in superfluids~\cite{Strocchi:2021abr}.

\item In Ref.~\cite{Lowdon:2022xcl} it was shown that pions in QCD persist above the pseudo-critical temperature, and have the structure of \textit{massive} thermoparticles. In the chiral limit this indicates that pions should behave like massless thermoparticle states above the critical temperature, which is precisely what one expects from Ref.~\cite{Bros:1996kd}, since pions are genuine Goldstone bosons in this limit\footnote{For a recent discussion of this issue see Ref.~\cite{Giordano:2022ghy}.}. These observations suggest that the high-temperature phase transitions in both chiral and physical QCD are not indicative of a change in the physical degrees of freedom associated with deconfinement, but may in fact be a reflection of the change in dissipative effects experienced by the (pseudo-)Goldstone bosons as the system increases in temperature. This is also supported by the Goldstone damping factor $D_{\beta}^{G}(\vec{x})$ in the $\mathrm{U}(1)$ scalar theory above $T_{c}$ having the same qualitative form as that observed for pions in Ref.~\cite{Lowdon:2022xcl}, even though the states in the scalar theory need not be composite.

\end{itemize}

\noindent
{\bf{\textit{Conclusions}}} \\
We have explored the fundamental question of how spontaneously broken symmetries manifest themselves at non-zero temperature, and in particular what happens to the Goldstone bosons that exist in the vacuum theory as the temperature is increased. In Ref.~\cite{Bros:1996kd} the authors established the non-perturbative signatures of thermal Goldstone bosons, and found that they need not cease to exist, even above the critical temperature $T_{c}$ where the symmetry is restored. In this work, we investigated $\mathrm{U}(1)$ scalar field theory on the lattice, analysing how the scalar correlation functions behave in the broken and high-temperature symmetry-restored phases of the theory. We demonstrated that the Goldstone boson does indeed continue to exist above $T_{c}$, and has the properties of a screened massless particle-like excitation, a so-called \textit{thermoparticle}. Since these characteristics are predicted to be model independent, they have implications for systems in which a continuous symmetry is restored at high temperatures. In particular, for zero or low-density non-relativistic systems this indicates that Goldstone modes should leave distinct signatures for $T>T_{c}$, and in QCD it suggests that the thermal evolution of pions around the chiral crossover may not be due to a fundamental change in the degrees of freedom, but could instead be driven by changes in the dissipative effects experienced by these states.

\begin{acknowledgments}
The authors acknowledge support by the Deutsche Forschungsgemeinschaft (DFG, German Research Foundation) through the Collaborative Research Center CRC-TR 211 ``Strong-interaction matter under extreme conditions'' -- Project No. 315477589-TRR 211. O.~P.~also acknowledges support by the State of Hesse within the Research Cluster ELEMENTS (Project ID 500/10.006).
\end{acknowledgments}

\bibliography{refs}

\begin{thebibliography}{29}%
\makeatletter
\providecommand \@ifxundefined [1]{%
 \@ifx{#1\undefined}
}%
\providecommand \@ifnum [1]{%
 \ifnum #1\expandafter \@firstoftwo
 \else \expandafter \@secondoftwo
 \fi
}%
\providecommand \@ifx [1]{%
 \ifx #1\expandafter \@firstoftwo
 \else \expandafter \@secondoftwo
 \fi
}%
\providecommand \natexlab [1]{#1}%
\providecommand \enquote  [1]{``#1''}%
\providecommand \bibnamefont  [1]{#1}%
\providecommand \bibfnamefont [1]{#1}%
\providecommand \citenamefont [1]{#1}%
\providecommand \href@noop [0]{\@secondoftwo}%
\providecommand \href [0]{\begingroup \@sanitize@url \@href}%
\providecommand \@href[1]{\@@startlink{#1}\@@href}%
\providecommand \@@href[1]{\endgroup#1\@@endlink}%
\providecommand \@sanitize@url [0]{\catcode `\\12\catcode `\$12\catcode
  `\&12\catcode `\#12\catcode `\^12\catcode `\_12\catcode `\%12\relax}%
\providecommand \@@startlink[1]{}%
\providecommand \@@endlink[0]{}%
\providecommand \url  [0]{\begingroup\@sanitize@url \@url }%
\providecommand \@url [1]{\endgroup\@href {#1}{\urlprefix }}%
\providecommand \urlprefix  [0]{URL }%
\providecommand \Eprint [0]{\href }%
\providecommand \doibase [0]{https://doi.org/}%
\providecommand \selectlanguage [0]{\@gobble}%
\providecommand \bibinfo  [0]{\@secondoftwo}%
\providecommand \bibfield  [0]{\@secondoftwo}%
\providecommand \translation [1]{[#1]}%
\providecommand \BibitemOpen [0]{}%
\providecommand \bibitemStop [0]{}%
\providecommand \bibitemNoStop [0]{.\EOS\space}%
\providecommand \EOS [0]{\spacefactor3000\relax}%
\providecommand \BibitemShut  [1]{\csname bibitem#1\endcsname}%
\let\auto@bib@innerbib\@empty
\bibitem [{\citenamefont {Goldstone}\ \emph {et~al.}(1962)\citenamefont
  {Goldstone}, \citenamefont {Salam},\ and\ \citenamefont
  {Weinberg}}]{Goldstone:1962es}%
  \BibitemOpen
  \bibfield  {author} {\bibinfo {author} {\bibfnamefont {J.}~\bibnamefont
  {Goldstone}}, \bibinfo {author} {\bibfnamefont {A.}~\bibnamefont {Salam}},\
  and\ \bibinfo {author} {\bibfnamefont {S.}~\bibnamefont {Weinberg}},\
  }\bibfield  {title} {\bibinfo {title} {{Broken Symmetries}},\ }\href
  {https://doi.org/10.1103/PhysRev.127.965} {\bibfield  {journal} {\bibinfo
  {journal} {Phys. Rev.}\ }\textbf {\bibinfo {volume} {127}},\ \bibinfo {pages}
  {965} (\bibinfo {year} {1962})}\BibitemShut {NoStop}%
\bibitem [{\citenamefont {Kastler}\ \emph {et~al.}(1966)\citenamefont
  {Kastler}, \citenamefont {Robinson},\ and\ \citenamefont
  {Swieca}}]{Kastler:1966wdu}%
  \BibitemOpen
  \bibfield  {author} {\bibinfo {author} {\bibfnamefont {D.}~\bibnamefont
  {Kastler}}, \bibinfo {author} {\bibfnamefont {D.~W.}\ \bibnamefont
  {Robinson}},\ and\ \bibinfo {author} {\bibfnamefont {A.}~\bibnamefont
  {Swieca}},\ }\bibfield  {title} {\bibinfo {title} {{Conserved currents and
  associated symmetries; Goldstone's theorem}},\ }\href
  {https://doi.org/10.1007/BF01773346} {\bibfield  {journal} {\bibinfo
  {journal} {Commun. Math. Phys.}\ }\textbf {\bibinfo {volume} {2}},\ \bibinfo
  {pages} {108} (\bibinfo {year} {1966})}\BibitemShut {NoStop}%
\bibitem [{\citenamefont {Kirzhnits}\ and\ \citenamefont
  {Linde}(1972)}]{Kirzhnits:1972ut}%
  \BibitemOpen
  \bibfield  {author} {\bibinfo {author} {\bibfnamefont {D.~A.}\ \bibnamefont
  {Kirzhnits}}\ and\ \bibinfo {author} {\bibfnamefont {A.~D.}\ \bibnamefont
  {Linde}},\ }\bibfield  {title} {\bibinfo {title} {{Macroscopic Consequences
  of the Weinberg Model}},\ }\href
  {https://doi.org/10.1016/0370-2693(72)90109-8} {\bibfield  {journal}
  {\bibinfo  {journal} {Phys. Lett. B}\ }\textbf {\bibinfo {volume} {42}},\
  \bibinfo {pages} {471} (\bibinfo {year} {1972})}\BibitemShut {NoStop}%
\bibitem [{\citenamefont {Dolan}\ and\ \citenamefont
  {Jackiw}(1974)}]{Dolan:1973qd}%
  \BibitemOpen
  \bibfield  {author} {\bibinfo {author} {\bibfnamefont {L.}~\bibnamefont
  {Dolan}}\ and\ \bibinfo {author} {\bibfnamefont {R.}~\bibnamefont {Jackiw}},\
  }\bibfield  {title} {\bibinfo {title} {{Symmetry Behavior at Finite
  Temperature}},\ }\href {https://doi.org/10.1103/PhysRevD.9.3320} {\bibfield
  {journal} {\bibinfo  {journal} {Phys. Rev. D}\ }\textbf {\bibinfo {volume}
  {9}},\ \bibinfo {pages} {3320} (\bibinfo {year} {1974})}\BibitemShut
  {NoStop}%
\bibitem [{\citenamefont {Weinberg}(1974)}]{Weinberg:1974hy}%
  \BibitemOpen
  \bibfield  {author} {\bibinfo {author} {\bibfnamefont {S.}~\bibnamefont
  {Weinberg}},\ }\bibfield  {title} {\bibinfo {title} {{Gauge and Global
  Symmetries at High Temperature}},\ }\href
  {https://doi.org/10.1103/PhysRevD.9.3357} {\bibfield  {journal} {\bibinfo
  {journal} {Phys. Rev. D}\ }\textbf {\bibinfo {volume} {9}},\ \bibinfo {pages}
  {3357} (\bibinfo {year} {1974})}\BibitemShut {NoStop}%
\bibitem [{\citenamefont {Strocchi}(2021)}]{Strocchi:2021abr}%
  \BibitemOpen
  \bibfield  {author} {\bibinfo {author} {\bibfnamefont {F.}~\bibnamefont
  {Strocchi}},\ }\href {https://doi.org/10.1007/978-3-662-62166-0} {\emph
  {\bibinfo {title} {{Symmetry Breaking}}}}\ (\bibinfo  {publisher}
  {Springer},\ \bibinfo {year} {2021})\BibitemShut {NoStop}%
\bibitem [{\citenamefont {Bros}\ and\ \citenamefont
  {Buchholz}(1998)}]{Bros:1996kd}%
  \BibitemOpen
  \bibfield  {author} {\bibinfo {author} {\bibfnamefont {J.}~\bibnamefont
  {Bros}}\ and\ \bibinfo {author} {\bibfnamefont {D.}~\bibnamefont
  {Buchholz}},\ }\bibfield  {title} {\bibinfo {title} {{The Unmasking of
  thermal Goldstone bosons}},\ }\href
  {https://doi.org/10.1103/PhysRevD.58.125012} {\bibfield  {journal} {\bibinfo
  {journal} {Phys. Rev. D}\ }\textbf {\bibinfo {volume} {58}},\ \bibinfo
  {pages} {125012} (\bibinfo {year} {1998})},\ \Eprint
  {https://arxiv.org/abs/hep-th/9608139} {arXiv:hep-th/9608139} \BibitemShut
  {NoStop}%
\bibitem [{\citenamefont {Bros}\ and\ \citenamefont
  {Buchholz}(1996)}]{Bros:1996mw}%
  \BibitemOpen
  \bibfield  {author} {\bibinfo {author} {\bibfnamefont {J.}~\bibnamefont
  {Bros}}\ and\ \bibinfo {author} {\bibfnamefont {D.}~\bibnamefont
  {Buchholz}},\ }\bibfield  {title} {\bibinfo {title} {{Axiomatic analyticity
  properties and representations of particles in thermal quantum field
  theory}},\ }\href@noop {} {\bibfield  {journal} {\bibinfo  {journal} {Ann.
  Inst. H. Poincare Phys. Theor.}\ }\textbf {\bibinfo {volume} {64}},\ \bibinfo
  {pages} {495} (\bibinfo {year} {1996})},\ \Eprint
  {https://arxiv.org/abs/hep-th/9606046} {arXiv:hep-th/9606046} \BibitemShut
  {NoStop}%
\bibitem [{\citenamefont {Streater}\ and\ \citenamefont
  {Wightman}(1989)}]{Streater:1989vi}%
  \BibitemOpen
  \bibfield  {author} {\bibinfo {author} {\bibfnamefont {R.~F.}\ \bibnamefont
  {Streater}}\ and\ \bibinfo {author} {\bibfnamefont {A.~S.}\ \bibnamefont
  {Wightman}},\ }\href@noop {} {\emph {\bibinfo {title} {{PCT, spin and
  statistics, and all that}}}}\ (\bibinfo  {publisher} {Redwood City, USA:
  Addison-Wesley},\ \bibinfo {year} {1989})\BibitemShut {NoStop}%
\bibitem [{\citenamefont {Haag}(1992)}]{Haag:1992hx}%
  \BibitemOpen
  \bibfield  {author} {\bibinfo {author} {\bibfnamefont {R.}~\bibnamefont
  {Haag}},\ }\href@noop {} {\emph {\bibinfo {title} {{Local quantum physics:
  Fields, particles, algebras}}}}\ (\bibinfo  {publisher} {Berlin, Germany:
  Springer},\ \bibinfo {year} {1992})\BibitemShut {NoStop}%
\bibitem [{\citenamefont {Bogolyubov}\ \emph {et~al.}(1990)\citenamefont
  {Bogolyubov}, \citenamefont {Logunov}, \citenamefont {Oksak},\ and\
  \citenamefont {Todorov}}]{Bogolyubov:1990kw}%
  \BibitemOpen
  \bibfield  {author} {\bibinfo {author} {\bibfnamefont {N.~N.}\ \bibnamefont
  {Bogolyubov}}, \bibinfo {author} {\bibfnamefont {A.~A.}\ \bibnamefont
  {Logunov}}, \bibinfo {author} {\bibfnamefont {A.~I.}\ \bibnamefont {Oksak}},\
  and\ \bibinfo {author} {\bibfnamefont {I.~T.}\ \bibnamefont {Todorov}},\
  }\href@noop {} {\emph {\bibinfo {title} {{General Principles of Quantum Field
  Theory}}}}\ (\bibinfo  {publisher} {Dordrecht, Netherlands: Kluwer},\
  \bibinfo {year} {1990})\BibitemShut {NoStop}%
\bibitem [{\citenamefont {Bros}\ and\ \citenamefont
  {Buchholz}(1992)}]{Bros:1992ey}%
  \BibitemOpen
  \bibfield  {author} {\bibinfo {author} {\bibfnamefont {J.}~\bibnamefont
  {Bros}}\ and\ \bibinfo {author} {\bibfnamefont {D.}~\bibnamefont
  {Buchholz}},\ }\bibfield  {title} {\bibinfo {title} {{Particles and
  propagators in relativistic thermo field theory}},\ }\href
  {https://doi.org/10.1007/BF01565114} {\bibfield  {journal} {\bibinfo
  {journal} {Z. Phys. C}\ }\textbf {\bibinfo {volume} {55}},\ \bibinfo {pages}
  {509} (\bibinfo {year} {1992})}\BibitemShut {NoStop}%
\bibitem [{\citenamefont {{K\"{a}ll\'{e}n}}(1952)}]{Kallen:1952zz}%
  \BibitemOpen
  \bibfield  {author} {\bibinfo {author} {\bibfnamefont {G.}~\bibnamefont
  {{K\"{a}ll\'{e}n}}},\ }\bibfield  {title} {\bibinfo {title} {{On the
  definition of the Renormalization Constants in Quantum Electrodynamics}},\
  }\href@noop {} {\bibfield  {journal} {\bibinfo  {journal} {Helv. Phys. Acta}\
  }\textbf {\bibinfo {volume} {25}},\ \bibinfo {pages} {417} (\bibinfo {year}
  {1952})}\BibitemShut {NoStop}%
\bibitem [{\citenamefont {Lehmann}(1954)}]{Lehmann:1954xi}%
  \BibitemOpen
  \bibfield  {author} {\bibinfo {author} {\bibfnamefont {H.}~\bibnamefont
  {Lehmann}},\ }\bibfield  {title} {\bibinfo {title} {{On the Properties of
  propagation functions and renormalization contants of quantized fields}},\
  }\href {https://doi.org/10.1007/BF02783624} {\bibfield  {journal} {\bibinfo
  {journal} {Nuovo Cim.}\ }\textbf {\bibinfo {volume} {11}},\ \bibinfo {pages}
  {342} (\bibinfo {year} {1954})}\BibitemShut {NoStop}%
\bibitem [{\citenamefont {Buchholz}(1993)}]{Buchholz:1993kp}%
  \BibitemOpen
  \bibfield  {author} {\bibinfo {author} {\bibfnamefont {D.}~\bibnamefont
  {Buchholz}},\ }\bibfield  {title} {\bibinfo {title} {{On the manifestations
  of particles}},\ }in\ \href@noop {} {\emph {\bibinfo {booktitle}
  {{International Conference on Mathematical Physics Towards the 21st
  Century}}}}\ (\bibinfo {year} {1993})\ \Eprint
  {https://arxiv.org/abs/hep-th/9511023} {arXiv:hep-th/9511023} \BibitemShut
  {NoStop}%
\bibitem [{\citenamefont {Lowdon}\ and\ \citenamefont
  {Philipsen}(2024)}]{Lowdon:2024atn}%
  \BibitemOpen
  \bibfield  {author} {\bibinfo {author} {\bibfnamefont {P.}~\bibnamefont
  {Lowdon}}\ and\ \bibinfo {author} {\bibfnamefont {O.}~\bibnamefont
  {Philipsen}},\ }\bibfield  {title} {\bibinfo {title} {{On the (in)consistency
  of perturbation theory at finite temperature}},\ }\href
  {https://doi.org/10.1007/JHEP08(2024)167} {\bibfield  {journal} {\bibinfo
  {journal} {JHEP}\ }\textbf {\bibinfo {volume} {08}},\ \bibinfo {pages}
  {167}},\ \Eprint {https://arxiv.org/abs/2405.02009} {arXiv:2405.02009
  [hep-ph]} \BibitemShut {NoStop}%
\bibitem [{\citenamefont {Lowdon}\ and\ \citenamefont
  {Philipsen}(2022)}]{Lowdon:2022xcl}%
  \BibitemOpen
  \bibfield  {author} {\bibinfo {author} {\bibfnamefont {P.}~\bibnamefont
  {Lowdon}}\ and\ \bibinfo {author} {\bibfnamefont {O.}~\bibnamefont
  {Philipsen}},\ }\bibfield  {title} {\bibinfo {title} {{Pion spectral
  properties above the chiral crossover of QCD}},\ }\href
  {https://doi.org/10.1007/JHEP10(2022)161} {\bibfield  {journal} {\bibinfo
  {journal} {JHEP}\ }\textbf {\bibinfo {volume} {10}},\ \bibinfo {pages}
  {161}},\ \Eprint {https://arxiv.org/abs/2207.14718} {arXiv:2207.14718
  [hep-lat]} \BibitemShut {NoStop}%
\bibitem [{\citenamefont {Bala}\ \emph {et~al.}(2024)\citenamefont {Bala},
  \citenamefont {Kaczmarek}, \citenamefont {Lowdon}, \citenamefont
  {Philipsen},\ and\ \citenamefont {Ueding}}]{Bala:2023iqu}%
  \BibitemOpen
  \bibfield  {author} {\bibinfo {author} {\bibfnamefont {D.}~\bibnamefont
  {Bala}}, \bibinfo {author} {\bibfnamefont {O.}~\bibnamefont {Kaczmarek}},
  \bibinfo {author} {\bibfnamefont {P.}~\bibnamefont {Lowdon}}, \bibinfo
  {author} {\bibfnamefont {O.}~\bibnamefont {Philipsen}},\ and\ \bibinfo
  {author} {\bibfnamefont {T.}~\bibnamefont {Ueding}},\ }\bibfield  {title}
  {\bibinfo {title} {{Pseudo-scalar meson spectral properties in the chiral
  crossover region of QCD}},\ }\href {https://doi.org/10.1007/JHEP05(2024)332}
  {\bibfield  {journal} {\bibinfo  {journal} {JHEP}\ }\textbf {\bibinfo
  {volume} {05}},\ \bibinfo {pages} {332}},\ \Eprint
  {https://arxiv.org/abs/2310.13476} {arXiv:2310.13476 [hep-lat]} \BibitemShut
  {NoStop}%
\bibitem [{\citenamefont {Haag}\ \emph {et~al.}(1967)\citenamefont {Haag},
  \citenamefont {Hugenholtz},\ and\ \citenamefont {Winnink}}]{Haag:1967sg}%
  \BibitemOpen
  \bibfield  {author} {\bibinfo {author} {\bibfnamefont {R.}~\bibnamefont
  {Haag}}, \bibinfo {author} {\bibfnamefont {N.~M.}\ \bibnamefont
  {Hugenholtz}},\ and\ \bibinfo {author} {\bibfnamefont {M.}~\bibnamefont
  {Winnink}},\ }\bibfield  {title} {\bibinfo {title} {{On the Equilibrium
  states in quantum statistical mechanics}},\ }\href
  {https://doi.org/10.1007/BF01646342} {\bibfield  {journal} {\bibinfo
  {journal} {Commun. Math. Phys.}\ }\textbf {\bibinfo {volume} {5}},\ \bibinfo
  {pages} {215} (\bibinfo {year} {1967})}\BibitemShut {NoStop}%
\bibitem [{\citenamefont {Kapusta}\ and\ \citenamefont
  {Gale}(2011)}]{Kapusta:2006pm}%
  \BibitemOpen
  \bibfield  {author} {\bibinfo {author} {\bibfnamefont {J.~I.}\ \bibnamefont
  {Kapusta}}\ and\ \bibinfo {author} {\bibfnamefont {C.}~\bibnamefont {Gale}},\
  }\href@noop {} {\emph {\bibinfo {title} {{Finite-temperature Field Theory:
  Principles and applications}}}},\ Cambridge Monographs on Mathematical
  Physics\ (\bibinfo  {publisher} {Cambridge University Press},\ \bibinfo
  {year} {2011})\BibitemShut {NoStop}%
\bibitem [{\citenamefont {Pilaftsis}\ and\ \citenamefont
  {Teresi}(2013)}]{Pilaftsis:2013xna}%
  \BibitemOpen
  \bibfield  {author} {\bibinfo {author} {\bibfnamefont {A.}~\bibnamefont
  {Pilaftsis}}\ and\ \bibinfo {author} {\bibfnamefont {D.}~\bibnamefont
  {Teresi}},\ }\bibfield  {title} {\bibinfo {title} {{Symmetry Improved CJT
  Effective Action}},\ }\href {https://doi.org/10.1016/j.nuclphysb.2013.06.004}
  {\bibfield  {journal} {\bibinfo  {journal} {Nucl. Phys. B}\ }\textbf
  {\bibinfo {volume} {874}},\ \bibinfo {pages} {594} (\bibinfo {year}
  {2013})},\ \Eprint {https://arxiv.org/abs/1305.3221} {arXiv:1305.3221
  [hep-ph]} \BibitemShut {NoStop}%
\bibitem [{\citenamefont {Neuberger}(1988{\natexlab{a}})}]{Neuberger:1987fd}%
  \BibitemOpen
  \bibfield  {author} {\bibinfo {author} {\bibfnamefont {H.}~\bibnamefont
  {Neuberger}},\ }\bibfield  {title} {\bibinfo {title} {{Soft Pions in Large
  Boxes}},\ }\href {https://doi.org/10.1016/0550-3213(88)90592-5} {\bibfield
  {journal} {\bibinfo  {journal} {Nucl. Phys. B}\ }\textbf {\bibinfo {volume}
  {300}},\ \bibinfo {pages} {180} (\bibinfo {year}
  {1988}{\natexlab{a}})}\BibitemShut {NoStop}%
\bibitem [{\citenamefont {Neuberger}(1988{\natexlab{b}})}]{Neuberger:1987zz}%
  \BibitemOpen
  \bibfield  {author} {\bibinfo {author} {\bibfnamefont {H.}~\bibnamefont
  {Neuberger}},\ }\bibfield  {title} {\bibinfo {title} {{A Better Way to
  Measure F($\pi$) in the Linear $\sigma$ Model}},\ }\href
  {https://doi.org/10.1103/PhysRevLett.60.889} {\bibfield  {journal} {\bibinfo
  {journal} {Phys. Rev. Lett.}\ }\textbf {\bibinfo {volume} {60}},\ \bibinfo
  {pages} {889} (\bibinfo {year} {1988}{\natexlab{b}})}\BibitemShut {NoStop}%
\bibitem [{\citenamefont {Hasenfratz}\ \emph {et~al.}(1987)\citenamefont
  {Hasenfratz}, \citenamefont {Jansen}, \citenamefont {Lang}, \citenamefont
  {Neuhaus},\ and\ \citenamefont {Yoneyama}}]{Hasenfratz:1987eh}%
  \BibitemOpen
  \bibfield  {author} {\bibinfo {author} {\bibfnamefont {A.}~\bibnamefont
  {Hasenfratz}}, \bibinfo {author} {\bibfnamefont {K.}~\bibnamefont {Jansen}},
  \bibinfo {author} {\bibfnamefont {C.~B.}\ \bibnamefont {Lang}}, \bibinfo
  {author} {\bibfnamefont {T.}~\bibnamefont {Neuhaus}},\ and\ \bibinfo {author}
  {\bibfnamefont {H.}~\bibnamefont {Yoneyama}},\ }\bibfield  {title} {\bibinfo
  {title} {{The Triviality Bound of the Four Component phi**4 Model}},\ }\href
  {https://doi.org/10.1016/0370-2693(87)91622-4} {\bibfield  {journal}
  {\bibinfo  {journal} {Phys. Lett. B}\ }\textbf {\bibinfo {volume} {199}},\
  \bibinfo {pages} {531} (\bibinfo {year} {1987})}\BibitemShut {NoStop}%
\bibitem [{\citenamefont {Hasenfratz}\ \emph {et~al.}(1989)\citenamefont
  {Hasenfratz}, \citenamefont {Jansen}, \citenamefont {Jersak}, \citenamefont
  {Lang}, \citenamefont {Neuhaus},\ and\ \citenamefont
  {Yoneyama}}]{Hasenfratz:1988kr}%
  \BibitemOpen
  \bibfield  {author} {\bibinfo {author} {\bibfnamefont {A.}~\bibnamefont
  {Hasenfratz}}, \bibinfo {author} {\bibfnamefont {K.}~\bibnamefont {Jansen}},
  \bibinfo {author} {\bibfnamefont {J.}~\bibnamefont {Jersak}}, \bibinfo
  {author} {\bibfnamefont {C.~B.}\ \bibnamefont {Lang}}, \bibinfo {author}
  {\bibfnamefont {T.}~\bibnamefont {Neuhaus}},\ and\ \bibinfo {author}
  {\bibfnamefont {H.}~\bibnamefont {Yoneyama}},\ }\bibfield  {title} {\bibinfo
  {title} {{Study of the Four Component phi**4 Model}},\ }\href
  {https://doi.org/10.1016/0550-3213(89)90562-2} {\bibfield  {journal}
  {\bibinfo  {journal} {Nucl. Phys. B}\ }\textbf {\bibinfo {volume} {317}},\
  \bibinfo {pages} {81} (\bibinfo {year} {1989})}\BibitemShut {NoStop}%
\bibitem [{\citenamefont {Hasenfratz}\ \emph {et~al.}(1990)\citenamefont
  {Hasenfratz}, \citenamefont {Jansen}, \citenamefont {Jersak}, \citenamefont
  {Lang}, \citenamefont {Leutwyler},\ and\ \citenamefont
  {Neuhaus}}]{Hasenfratz:1989ux}%
  \BibitemOpen
  \bibfield  {author} {\bibinfo {author} {\bibfnamefont {A.}~\bibnamefont
  {Hasenfratz}}, \bibinfo {author} {\bibfnamefont {K.}~\bibnamefont {Jansen}},
  \bibinfo {author} {\bibfnamefont {J.}~\bibnamefont {Jersak}}, \bibinfo
  {author} {\bibfnamefont {C.~B.}\ \bibnamefont {Lang}}, \bibinfo {author}
  {\bibfnamefont {H.}~\bibnamefont {Leutwyler}},\ and\ \bibinfo {author}
  {\bibfnamefont {T.}~\bibnamefont {Neuhaus}},\ }\bibfield  {title} {\bibinfo
  {title} {{Finite Size Effects and Spontaneously Broken Symmetries: The Case
  of the $\mathrm{O}(4)$ Model}},\ }\href {https://doi.org/10.1007/BF01556001}
  {\bibfield  {journal} {\bibinfo  {journal} {Z. Phys. C}\ }\textbf {\bibinfo
  {volume} {46}},\ \bibinfo {pages} {257} (\bibinfo {year} {1990})}\BibitemShut
  {NoStop}%
\bibitem [{\citenamefont {Hasenfratz}\ \emph {et~al.}(1991)\citenamefont
  {Hasenfratz}, \citenamefont {Jansen}, \citenamefont {Jersak}, \citenamefont
  {Kastrup}, \citenamefont {Lang}, \citenamefont {Leutwyler},\ and\
  \citenamefont {Neuhaus}}]{Hasenfratz:1990fu}%
  \BibitemOpen
  \bibfield  {author} {\bibinfo {author} {\bibfnamefont {A.}~\bibnamefont
  {Hasenfratz}}, \bibinfo {author} {\bibfnamefont {K.}~\bibnamefont {Jansen}},
  \bibinfo {author} {\bibfnamefont {J.}~\bibnamefont {Jersak}}, \bibinfo
  {author} {\bibfnamefont {H.~A.}\ \bibnamefont {Kastrup}}, \bibinfo {author}
  {\bibfnamefont {C.~B.}\ \bibnamefont {Lang}}, \bibinfo {author}
  {\bibfnamefont {H.}~\bibnamefont {Leutwyler}},\ and\ \bibinfo {author}
  {\bibfnamefont {T.}~\bibnamefont {Neuhaus}},\ }\bibfield  {title} {\bibinfo
  {title} {{Goldstone bosons and finite size effects: A Numerical study of the
  $\mathrm{O}(4)$ model}},\ }\href
  {https://doi.org/10.1016/0550-3213(91)90153-O} {\bibfield  {journal}
  {\bibinfo  {journal} {Nucl. Phys. B}\ }\textbf {\bibinfo {volume} {356}},\
  \bibinfo {pages} {332} (\bibinfo {year} {1991})}\BibitemShut {NoStop}%
\bibitem [{\citenamefont {Gockeler}\ \emph {et~al.}(1993)\citenamefont
  {Gockeler}, \citenamefont {Kastrup}, \citenamefont {Neuhaus},\ and\
  \citenamefont {Zimmermann}}]{Gockeler:1992zj}%
  \BibitemOpen
  \bibfield  {author} {\bibinfo {author} {\bibfnamefont {M.}~\bibnamefont
  {Gockeler}}, \bibinfo {author} {\bibfnamefont {H.~A.}\ \bibnamefont
  {Kastrup}}, \bibinfo {author} {\bibfnamefont {T.}~\bibnamefont {Neuhaus}},\
  and\ \bibinfo {author} {\bibfnamefont {F.}~\bibnamefont {Zimmermann}},\
  }\bibfield  {title} {\bibinfo {title} {{Scaling analysis of the
  $\mathrm{O}(4)$ symmetric Phi**4 theory in the broken phase}},\ }\href
  {https://doi.org/10.1016/0550-3213(93)90489-C} {\bibfield  {journal}
  {\bibinfo  {journal} {Nucl. Phys. B}\ }\textbf {\bibinfo {volume} {404}},\
  \bibinfo {pages} {517} (\bibinfo {year} {1993})},\ \Eprint
  {https://arxiv.org/abs/hep-lat/9206025} {arXiv:hep-lat/9206025} \BibitemShut
  {NoStop}%
\bibitem [{\citenamefont {Giordano}(2022)}]{Giordano:2022ghy}%
  \BibitemOpen
  \bibfield  {author} {\bibinfo {author} {\bibfnamefont {M.}~\bibnamefont
  {Giordano}},\ }\bibfield  {title} {\bibinfo {title} {{Localisation of Dirac
  modes in gauge theories and Goldstone\textquoteright{}s theorem at finite
  temperature}},\ }\href {https://doi.org/10.1007/JHEP12(2022)103} {\bibfield
  {journal} {\bibinfo  {journal} {JHEP}\ }\textbf {\bibinfo {volume} {12}},\
  \bibinfo {pages} {103}},\ \Eprint {https://arxiv.org/abs/2206.11109}
  {arXiv:2206.11109 [hep-th]} \BibitemShut {NoStop}%
\end{thebibliography}%
   
\end{document}